\documentclass[sn-mathphys,Numbered,iicol,onecolumn]{sn-jnl}

\usepackage{graphicx}%
\usepackage{multirow}%
\usepackage{amsmath,amssymb,amsfonts}%
\usepackage{amsthm}%
\usepackage{mathrsfs}%
\usepackage[title]{appendix}%
\usepackage{xcolor}%
\usepackage{textcomp}%
\usepackage{manyfoot}%
\usepackage{booktabs}%
\usepackage{algorithm}%
\usepackage{algorithmicx}%
\usepackage{algpseudocode}%
\usepackage{listings}%

\theoremstyle{thmstyleone}%
\theoremstyle{thmstyletwo}%

\theoremstyle{thmstylethree}%

\begin{document}

\title[Article Title]{Characterizing the heterogeneity of membrane
  liquid-ordered domains}


\author[1]{\fnm{Tanmoy} \sur{Sarkar}}\email{tanmoy.sarkar@ijs.si}

\author*[2]{\fnm{Oded} \sur{Farago}}\email{ofarago@bgu.ac.il}


\affil[1]{\orgdiv{Department of Theoretical Physics}, \orgname{Jo\v zef Stefan Institute}, \orgaddress{\street{Jamova 39}, \city{Ljubljana}, \postcode{1000}, \country{ Slovenia}}}

\affil*[2]{\orgdiv{Biomedical Engineering Department}, \orgname{Ben Gurion University of the Negev}, \orgaddress{\city{Beer Sheva}, \postcode{84105}, \country{Israel}}}

\abstract{We use a lattice model of a ternary mixture containing
  saturated and unsaturated lipids with cholesterol (Chol), to study
  the structural properties characterizing the coexistence between the
  liquid-disordered and liquid-ordered phases. Depending on the
  affinity of the saturated and unsaturated lipids, the system may
  exhibit macroscopic (thermodynamic) liquid-liquid phase separation,
  or be divided into small-size liquid-ordered domains surrounded by a
  liquid-disordered matrix. In both cases, it is found that the
  nano-scale structure of the liquid-ordered regions is heterogeneous,
  and that they are partitioned into Chol-rich sub-domains and
  Chol-free, gel-like, nano-clusters. This emerges as a characteristic
  feature of the liquid-ordered state, which helps distinguishing
  between liquid-ordered domains in a two-phase mixture, and
  similar-looking domains in a one-phase mixture that are rich in
  saturated lipids and Chol, but are merely thermal density
  fluctuations. The nano-structure heterogeneity of the liquid-ordered
  phase can be detected by suitable experimental spectroscopic
  methods, and is observed also in atomistic computer simulations.}




\maketitle

\section{Introduction}\label{sec1}

The lateral heterogeneity of the plasma membrane (PM) is an intriguing
topic in membrane biophysics. The interest in this problem stems from
the widely-accepted {\em raft hypothesis}\/ that the PM contains small
(10-200 nm) and dynamic domains that are enriched in sphingolipids and
cholesterol (Chol), and associate with specific proteins while
excluding others \cite{simons97, pike06, levental20}. Raft domains
play a key role in cellular processes such as signal transduction
\cite{simons2000}, cell adhesion \cite{leitinger2002, murai2012}, and
membrane trafficking \cite{ikonen2001,hanzal2007}. Physically, raft
domains are in a liquid-ordered (Lo) state, which combines properties
of the two main phases of lipid bilayers - the liquid-disordered (Ld)
and gel (So) phases \cite{vdgoot01,kaiser09}. Similarly to the former,
the lipids in the liquid-ordered domains are mobile and free to
diffuse \cite{filippov04} in the membrane plane. However, their
hydrocarbon chains are ordered, fully extended and tightly packed, as
in the gel phase \cite{holl08}, which causes the liquid-ordered
domains to be more viscous than the liquid-disordered matrix where the
rafts float.

From a thermodynamic perspective, the fundamental question is whether
the presence of raft domains in the PM is an example of liquid-liquid
phase coexistence and, if so, why the two liquid phases do not
separate macroscopically in order to minimize their interfacial free
energy~\cite{lenne09}. Obviously, this puzzle pertains only to systems
that are at thermodynamics equilibrium, while the PM of a living cell
is an active system. Moreover, the PM is connected by various protein
complexes to both the extracellular matrix and the cell cytoskeleton,
which means that it cannot be considered as an isolated physical
system. Nevertheless, because of the enormous complexity of biological
membranes, much of the discussion on the problem of liquid-liquid
phase separation in the PM has been based on investigations of simple
model systems, especially ternary mixtures of saturated and
unsaturated lipids with Chol \cite{veatch05, veatch07, goni08,
  feigenson09, uppamoochikkal, hirst11, komura14, levental16}. These
studies reveal that for certain molar ratios and temperatures, such
ternary mixtures may separate into Ld and Lo regions, where the latter
are indeed enriched in saturated lipids and Chol. For some mixtures,
e.g., the ternary mixture containing DPPC (saturated), DOPC
(unsaturated), and Chol at $T=280$K, the two liquid phases are
macroscopically (thermodynamically) separated and micron-size
liquid-ordered domains can be observed in fluorescence microscope
images \cite{veatch03, veatch07, zhao07}. Others, e.g., the ternary
mixture DPPC/POPC/Chol, feature no macroscopic phase separation and
appear uniform when investigated by direct imaging
methods~\cite{zhao07b}. However, they do contain nanoscopic
liquid-ordered domains whose presence is detected by indirect
spectroscopic methods that are sensitive to details on the nanometric
scales \cite{ clarke, frazier, Heberle13, yasuda15}. This is often
referred to as ``microscopic'' or ``local'' phase separation, although
it is clear that this situation does not describe coexistence of
macroscopic phases in the thermodynamic sense \cite{dealmeida,
  heberle10, schmid17review, komura14}.

\begin{figure*}[t]
\centering \includegraphics[width=0.9\textwidth]{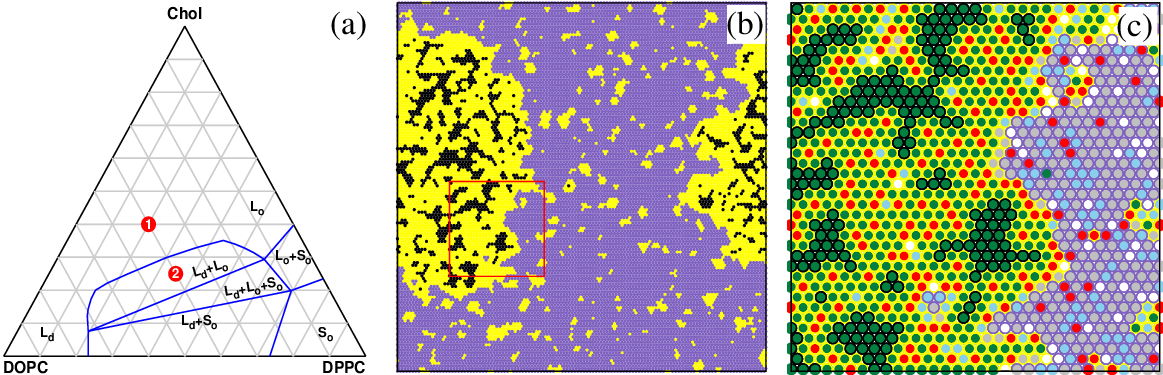}
\caption{(a) The phase diagram of the Type II ternary DPPC/DOPC/Chol
  mixture at $T=283$K. The blue phase separating lines are adapted
  from ref \cite{veatch07}. The red points indicate the compositions
  in the one- and two-phase regions where the simulations are
  conducted. (b) An equilibrium snapshot of the mixture at point (2)
  in the phase diagram for $\epsilon_{24}=0$. The liquid-disordered ,
  liquid-ordered, and gel sites are colored in purple, yellow, and
  black, respectively. (c) A magnification of the region inside the
  red box in (b). Lattice sites are color-coded by their content as
  follows: ordered DPPC (green), disordered DPPC (blue), DOPC (grey),
  Chol (red), or empty (white).  Purple, yellow and black borders are
  assigned to the sites from liquid-disordered, liquid-ordered, and
  gel regions, respectively. Fig. (b) and (c) are adapted from
  ref. \cite{sarkar2023}}
\label{fig1}
\end{figure*}

\begin{figure*}[t]
\centering \includegraphics[width=0.9\textwidth]{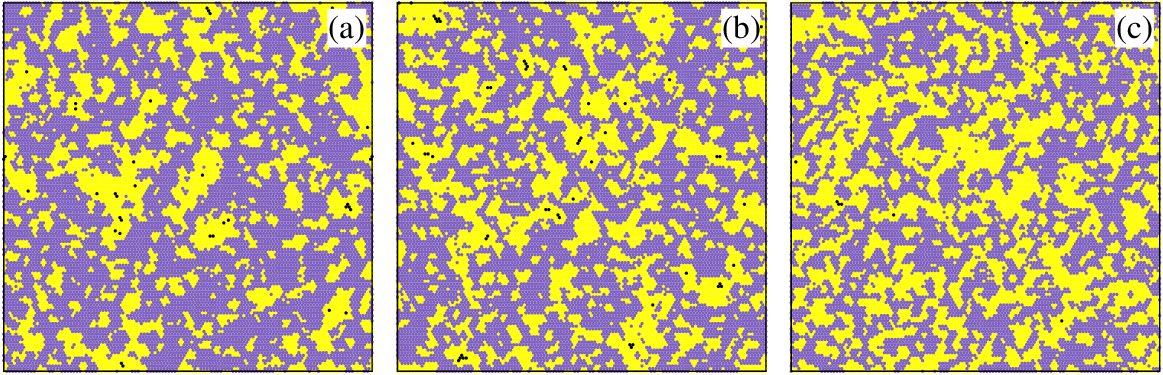}
\caption{(a) Equilibrium snapshots of the mixture at point (1) in the
  phase diagram [{\protect see Fig.~\ref{fig1}}(a)] for (a)
  $\epsilon_{24}=0$, (b) $\epsilon_{24}=0.2$, and (c)
  $\epsilon_{24}=0.4$. Color coding is as in {\protect
    Fig.~\ref{fig1}(b)}.}
\label{fig2}
\end{figure*}

Mixtures exhibiting macroscopic and microscopic phase separation are
commonly referred to as Type II and Type I mixtures, respectively
\cite{feigenson09}. Quite remarkably, the phase diagrams of many
ternary mixtures of both types look quite similar to the
DPPC/DOPC/Chol phase diagram depicted \cite{veatch07, veatch03, davis,
  uppamoochikkal, sarkar2023} in Fig.~\ref{fig1}(a) (with phase
boundaries that are obviously specific for each mixture). In Type II
mixtures, a macroscopic Lo phase is detected in the two-phase region
[see Fig.~\ref{fig1}(b) corresponding to point (2) in
  Fig.~\ref{fig1}(a)], while smaller liquid-ordered domains [see
  Fig.~\ref{fig2}(a) corresponding to point (1) in Fig.~\ref{fig1}(a)]
are found in the one-phase region. These nanoscopic domains represent
density fluctuations that are expected in the one-phase region near a
second-order critical point \cite{veatch07}. In Type I mixtures, the
physics behind the formation of such domains is less clear. As noted
above, the identification of the Ld-Lo coexistence region in the phase
diagram of Type I mixtures is based on the interpretation of
scattering data. This implies that certain nano-scale structural
changes take place when the system is taken from the one-phase to the
two-phase region. Indeed, experimental and computational studies
\cite{amstrong13, sodt, javanainen17, tieleman20, sarkar2021} have
shown that liquid-ordered domains contain clusters of tightly-packed
hexagonally-arranged acyl chains of the saturated lipids [see
  Fig.~\ref{fig1}(c)]. The existence of such gel-like clusters in
liquid-ordered domains has been termed ``nano-domains within
domains''. Notice that Chol is depleted from the gel-like
nano-cluster, which means that the distribution of components in the
liquid-ordered domains is spatially heterogeneous. The nano-scale
separation of the domains into Chol-rich and Chol-free regions is
consistent with the following thermodynamic explanation for the
micro-phase separation of Type I mixtures: Formation of many small
domains rather than macroscopic phase separation may indicate that the
line tension between the coexisting phases is vanishingly small. Since
the line tension between the Ld and Lo phases is primarily due to the
unfavorable mixing of the ordered chain of the saturated lipids and
the disordered chains of the unsaturated ones, the Chol may act as a
line active agent (lineactant), similarly to the way that surfactant
molecules eliminate the surface tension between oil and water and
allow for the formation of microemulsions~\cite{engberg16}. This
property follows from the {\em duality of the
  cholesterol}~\cite{berkowitz09}: On the one hand, it has affinity to
the saturated lipids which is the reason why it is largely depleted
from the Ld phase, but on the other hand, it disrupts the tight
packing of the saturated chains with each other, which may explain the
formation of the Chol-free gel-like clusters. In this picture, the
Chol accumulates in buffer regions between the gel-like and
liquid-disordered regions of the mixture which, we remind, are the
main phases of bilayer membranes that do not contain Chol.

Recently, we presented a minimal lattice model that captures correctly
the behavior of the ternary mixture DPPC/DOPC/Chol at both the
microscopic and macroscopic scales \cite{sarkar2023}. On the
macroscopic scale, the simulation results of mixtures with different
molar compositions showed very good agreement with the
experimentally-derived phase diagram of the mixture. On the
microscopic scales, the model produced agreement with atomistic
simulation results showing the formation of gel-like clusters inside
the liquid-ordered domains. By changing the value of one of the model
parameters that controls the affinity between the saturated and
unsaturated lipids, we demonstrated that it is possible to change the
nature of the ternary mixture from Type II to Type I, while preserving
the inhomogeneous distribution of the saturated lipids and Chol inside
the liquid-ordered domains. One of the interesting findings of our
previous study of the Type II DPPC/DOPC/Chol mixture was the fact that
while near-critical density fluctuations were indeed observed in the
one-phase regime of the system, the gel-like clusters within the
liquid-ordered regions appeared only in the two-phase coexistence
regime. Remarkably, this was also the case in the Type I counterpart
of the DPPC/DOPC/Chol mixture, i.e., the one obtained by a variation
of the model parameter associated with the mixing of saturated and
unsaturated lipids. The change in the model parameter caused the
system to exhibit micro- rather than macro-phase separation, but did
not cause the internal distribution of the lipid and Chol components
inside the liquid-ordered domains to become uniform. This raises the
possibility that the presence of such nano-clusters may be a defining
property of the liquid-ordered state. As these clusters can be
detected by some spectroscopic methods, their emergence can help
drawing the boundaries in the phase diagram of lipid mixtures.

Here, we quantitatively explore the internal structure of the Lo
state, both in Type II and Type I mixtures. For this purpose, we
extend our lattice simulations of ternary mixtures from
ref.~\cite{sarkar2023}, to study the multi-scale behavior of the
system as a function of the affinity between the saturated and
unsaturated lipids. Our study reveals that the Lo phase in Type II
mixtures is heterogeneous at the nanometric scale. When the affinity
of the saturated and unsaturated lipids becomes higher, the
macroscopic phase breaks into smaller domains but the internal
heterogeneity is preserved in those domains.  This property helps to
distinguish between small liquid-ordered domains (in Type I mixtures)
and density fluctuations of similar size occurring in the one-phase
region, where the fraction of saturated lipids is relatively low and
gel-like clusters are rarely observed. When the fraction of saturated
lipids is increased, the liquid-ordered domains become increasingly
populated with such clusters and gradually transform into the gel
state [see phase diagram, Fig.~\ref{fig1}(a)].

\bigskip

\section{Simulations and Results}

The details of the lattice model and the Monte Carlo (MC) simulations
are found in ref.~\cite{sarkar2023}. Briefly, we consider a triangular
lattice of $N_s=121\times 140=16940$ (having an aspect ratio which is
very close to unity) with periodic boundary conditions. The lattice is
populated by DPPC, DOPC, and Chol molecules, where the lipids are
modeled as dimers (i.e., each of their chains occupies a single site),
while Chol is represented as a monomer. Some lattice sites are left
empty, thereby allowing the molecules to diffuse on the lattice, which
mimics the fluidity of the liquid membrane. While the DOPC chains are
assumed to be disordered, the chains of the saturated DPPC lipids may
be either ordered or disordered. Thus, each lattice site can be in one
of five possible states: (i) empty ($s=0$), (ii) a disordered ($s=1$)
or (iii) an ordered ($s=2$) DPPC chain, (iv) a Chol ($s=3$), (v) a
DOPC chain ($s=4$). The energy of a configuration is given by
\begin{eqnarray}
  E&=&-\Omega_1 k_BT\sum_{i} \delta_{s_i,1}\nonumber \\
  &-&\sum_{i,j}\epsilon_{22}\delta_{s_i,2}\delta_{s_j,2}\nonumber \\
  &-&\sum_{i,j} \epsilon_{23}\left[\delta_{s_i,2}\delta_{s_j,3}
  +\delta_{s_i,3}\delta_{s_j,2}\right]\nonumber \\
  &-&\sum_{i,j} \epsilon_{24}\left[\delta_{s_i,2}\delta_{s_j,4}
    +\delta_{s_i,4}\delta_{s_j,2}\right],
  \label{eq:mcenergy}
\end{eqnarray}
where the deltas are Kronecker deltas, and the summations are carried
over all $N_s$ lattice sites in the first term and over all nearest
neighbor pair sites in the other terms.  The first term in
Eq.~(\ref{eq:mcenergy}) accounts for the fact that the disordered
state of the DPPC chains is entropically favored by a free energy,
$-\Omega_1 k_BT$ (where $k_B$ is Boltzmann constant and $T$ is the
temperature), over the ordered state. The other terms represent
effective packing attraction between an ordered DPPC chain with
another ordered DPPC chain (second term), Chol (third term), and DOPC
chains (fourth term). We set the model parameters to $\Omega_1=3.9$,
$\epsilon_{22}=1.3\epsilon$ and $\epsilon_{23}= 0.72\epsilon$, where
the energy unit $\epsilon$ is such that the melting temperature of a
pure DPPC membrane is $T_m=314{\rm K}=0.9\epsilon/k_B$ \cite{pan2008,
  sarkar2021}. For these values, we showed in ref.~\cite{sarkar2023}
that the model reproduces the experimentally-derived phase diagram for
$\epsilon_{24}=0$. In contrast, when we set $\epsilon=0.4$ while
keeping the other parameters unchanged, the macroscopic liquid-liquid
phase separation was lost, and the mixture changed its nature from
Type II to Type I. Here, we inspect how the configurational
characteristics of the system change as a function of the parameter
$\epsilon_{24}$, the one that controls the degree of mixing between
the saturated and unsaturated lipids. Interestingly, we find that
systems at compositions corresponding, respectively, to the one- and
two-phase regions of the Type II mixture phase diagram
($\epsilon_{24}=0$), retain different local characteristics when the
value of $\epsilon_{24}$ increases and the mixture transforms to Type
I. For this reason, we only draw the phase diagram of the Type II
mixture [Fig.~\ref{fig1}(a)], which also serves as a reference for the
simulations and discussion of the properties of Type I mixtures.

In order to identify the ordered and disordered regions in the system,
we define the following order parameter at each lattice site $i$ 
\begin{equation}
  G_i=Sc_i+\sum_{j=1}^6 Sc_j,
  \label{eq:grade}
\end{equation}
where $Sc_i$ receives one of the following values
$Sc_i=0,-0.5,2,1,-1$, if the state of a site is equal, respectively,
to $s=0,1,2,3,4$, and the sum in Eq.~\ref{eq:grade} is carried over
the six nearest-neighbor of the site (see ref.~\cite{sarkar2023} for
full details of this state-classification algorithm; we note that it
yields very good agreement with atomistic simulations
data~\cite{tieleman20} utilizing a different hidden Markov chain
algorithm for this purpose). The local order parameter $G_i$ takes
integer values between -7 and 14, where the sites with non-negative
(negative) value, $G_i\ge 0$ ($G_i<0$), are associated with the
liquid-ordered (liquid-disordered) regions. The maximum value of
$G_i=14$ is associated with the gel region. It is obtained at sites
where ordered DPPC chains are surrounded by 6 other ordered DPPC
chains. Fig.~\ref{fig2} shows equilibrium configurations of
DPPC/DOPC/Chol mixtures with mole fractions of 0.2/0.4/0.4 [point (1)
  in Fig.~\ref{fig1}(a)] for $\epsilon_{24}=0$ (a), 0.2 (b), and 0.4
(c). The snapshots show mixtures with many small liquid-ordered
domains surrounded by a liquid-disordered matrix. As noted earlier, in
Type II mixtures [snapshot (a)], the appearance of such domains may be
a precursor of a demixing phase transition. Interestingly, changing
the value of the model parameter $\epsilon_{24}$, does not modify the
general characteristics of the system. The type I mixture [snapshot
  (c)] exhibits very similar liquid-ordered domains.

\begin{figure}[t]
\centering \includegraphics[width=0.45\textwidth]{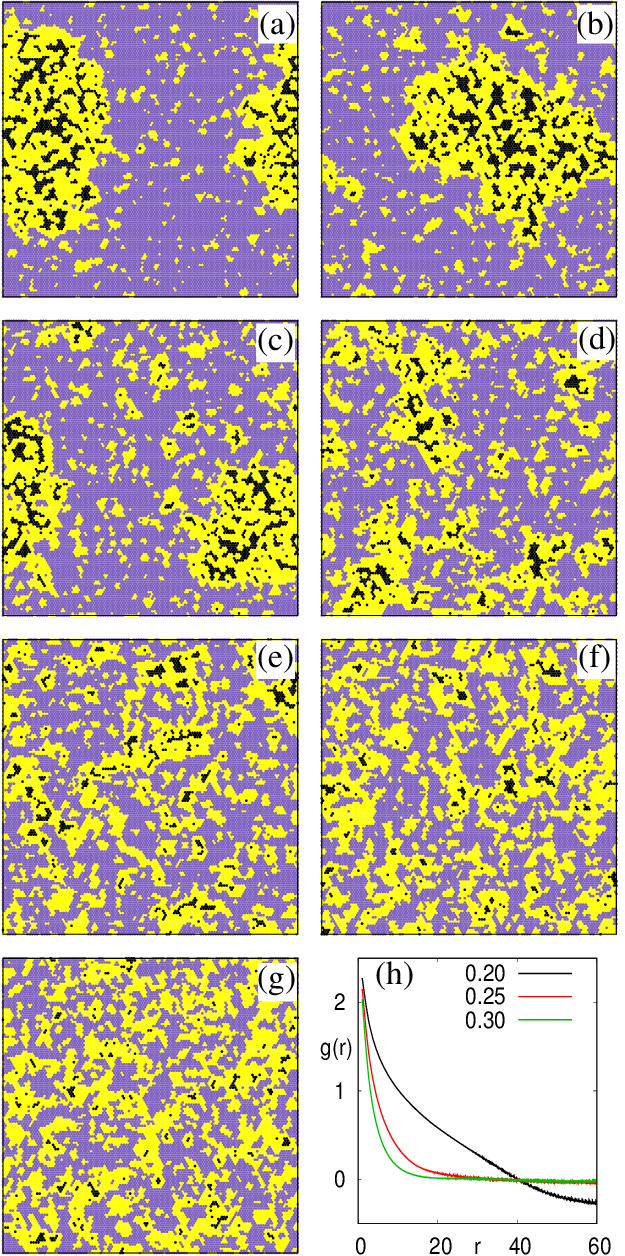}
\caption{Equilibrium snapshots of the mixture at point (2) in the
  phase diagram [{\protect see Fig.~\ref{fig1}}(a)]. (a)-(g) show,
  respctively, snapshots corresponding to $\epsilon_{24}=0$, 0.1, 0.2,
  0.25, 0.3, 0.35, and 0.4. Color coding is as in {\protect
    Fig.~\ref{fig1}(b)}. (h) The pair correlation function defined in
  Eq.~(\ref{eq:pair}) as a function of the pair distance $r$ for
  different values of $\epsilon_{24}$ (measured in units of the
  lattice spacing $l$).}
\label{fig3}
\end{figure}
\bigskip

Fig.~\ref{fig3} shows equilibrium configurations of DPPC/DOPC/Chol
mixtures with mole fractions of 0.35/0.4/0.25 [point (2) in
  Fig.~\ref{fig1}(a)] for $\epsilon_{24}=0$, 0.1, 0.2, 0.25, 0.3,
0.35, and 0.4 [snapshots (a)-(g), respectively]. This point lies in
the two-phase region, and the figure suggests that the mixture
undergoes a transition from Type II to Type I through a Lifshitz point
that is found between $\epsilon_{24}=0.2$ and
$\epsilon_{24}=0.25$. This is also evident from the calculation of the
pair correlation function, $g(r)$, defined by
\begin{equation}
  g(r)\equiv\frac{\left\langle G(x_i)G(x_i+r)\right\rangle}{\left\langle
    G(x_i)\right\rangle^2}-1,
  \label{eq:pair}
\end{equation}
where $r$ is the pair distance between lattice points, and the
averages are taken over all the lattice points $x_i$ and over 100
independent snapshots of the system. The correlation functions for
$\epsilon_{24}=0.2$ and $\epsilon_{24}=0.25$ are plotted in
Fig.~\ref{fig3}(h). In the former case, the behavior of $g(r)$ is
consistent with macroscopic Ld-Lo phase coexistence: The function
drops to zero at $r\sim40$, which is roughly a third of the size in
the simulated mixture, and then becomes negative at larger distances,
which is due to the fact that the order parameter assumes positive and
negative values in, respectively, the ordered and disordered regions
of the system. In the latter case, the function vanishes at all
distances larger than $r\sim 20$. This is consistent with the picture
that on scales larger than the domain size, the system is essentially
homogeneous. Calculation of $g(r)$ at a larger value
$\epsilon_{24}=0.3$ [see also Fig.~\ref{fig3}(h)] shows that the
characteristic domain size decreases as we move further away from the
point of transition from Type II to Type I.

\begin{figure*}[t]
\centering \includegraphics[width=0.9\textwidth]{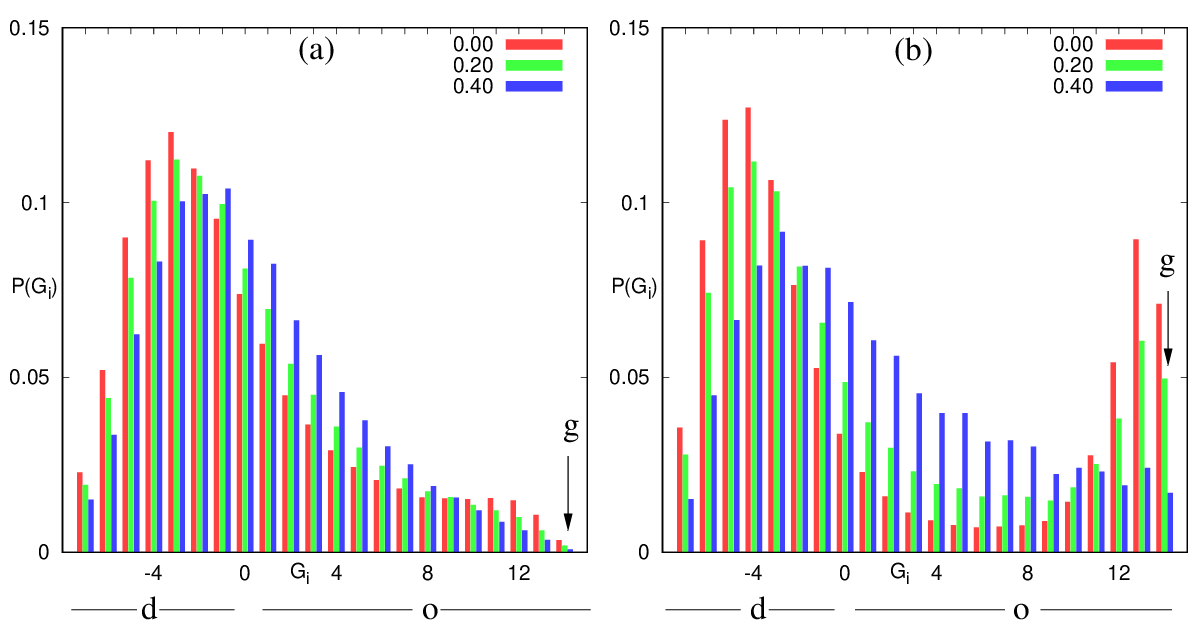}
\caption{(a) The distribution histograms of the values of the order
  parameter $G_i$ for mixtures with molar compositions indicated by
  point (1) in {\protect Fig.~\ref{fig1}}(a) (one-phase region) and
  for different values of the model parameter $\epsilon_{24}$. (b)
  Same as (a) for mixtures with molar compositions indicated by point
  (2) in {\protect Fig.~\ref{fig1}}(a) (two-phase region). The letters
  d and o below the figures mark the portions of the histograms
  corresponding to the disordered ($G_i<0$) and ordered ($G_i\geq0$)
  regions of the mixtures, respectively, while the letter g marks the
  last column ($G_i=14$) corresponding to the gel-like regions.}
\label{fig4}
\end{figure*}
\bigskip

Fig.~\ref{fig2} (a), (b), (c) and Fig.~\ref{fig3} (a), (c), (g) show
equilibrium snapshots of mixtures corresponding to $\epsilon=0$, 0.2,
and 0.4 in the one- and two-phase regions, respectively. Comparison of
these figures raises two major questions: (1) Are there any
configurational differences between the equilibrium snapshots in the
one- and two-phase regions of the Type I mixture
($\epsilon_{24}=0.4$)? (ii) Are there any configurational similarities
between the snapshots in the two-phase region in both Types II and
Type I mixtures ($\epsilon_{24}=0$ and 0.2
vs.~$\epsilon_{24}=0.4$)? The answers to these questions is found in
Fig.~\ref{fig4} showing the distribution histograms of the values of
the order parameter $G_i$ for these values of $\epsilon$. The
histograms corresponding to mixtures in the one- and two-phase regions
are plotted in (a) and (b), respectively. The differences between them
are clear: In the one-phase region, the distribution of order
parameters is unimodal, with a nearly-negligible fraction of highly
ordered gel-like sites ($G_i=14$). In contrast, the distribution in the
two-phase region is bimodal with a substantial fraction of gel-like
regions. Importantly, the bimodal distribution is found for both Type
II and Type I mixtures, implying that this is a characteristic feature
of the Lo state, irrespective of whether it appears in the form of a
macroscopic phase or small-size domains. Specifically to Type I
mixtures, our study reveals the fundamental difference between the
similar-looking figures~\ref{fig2}(c) and \ref{fig3}(g), thus
providing insight into the concept of microscopic liquid-liquid phase
separation. In the former figure, the ordered regions (shown in
yellow) are thermal density fluctuations in a one-phase system which
is characterized by a unimodal distribution of the order parameter. In
contrast, the distribution of local order parameter in the latter
figure is bimodal, which indicates that we are looking at two
different phases that are not macroscopically separated, presumably
because of the smallness of the line tension between them. The
emergence of a second peak in fig.~\ref{fig4}(b) at large values of
$G_i$ is directly related to the appearance of nanoscopic gel-like
clusters. These highly ordered regions serve as a characteristic
feature of the Lo phase, and their detection in computer simulations
or via suitable experimental methods helps identify the liquid-liquid
two-phase region of ternary mixture.  

\section{Conclusion}

We use a simple lattice model of ternary mixtures containing saturated
and unsaturated lipids with Chol, to quantitatively characterize the
liquid-ordered state observed in such mixtures at certain compositions
and temperatures. Contrary to the common perception that the
liquid-ordered domains are simply regions containing high
concentrations of saturated lipids and Chol, we find that these
domains are internally heterogeneous. Explicitly, the liquid-ordered
regions in ternary mixture are indeed enriched in saturated lipids and
Chol, but the distribution of these components on the nanometric scale
is not uniform. The statistics of the local order parameter, $G_i$,
emerges as a criterion for the distinction between one-phase mixtures
exhibiting thermal fluctuation resembling liquid-ordered domains, and
two-phase mixtures featuring liquid-liquid coexistence. In the former
case, the distribution is unimodal, while in the latter it is bimodal
and shows a second peak at large values of $G_i$. The second peak in
the distribution is due to the partition of the liquid-ordered domains
into Chol-rich and Chol-free gel-like sub-domains, and it appears in
both types of mixtures, i.e., either if the Ld and Lo phases are
separated macroscopically or microscopically. We hypothesize that the
heterogeneity of the Lo phase is related to the relative thermodynamic
stability of the gel (So) phase, where the saturated lipids are
tightly packed to each other, causing the Chol to be expelled from
these regions. In this picture, the Chol-rich regions in the Lo phase
may be regarded as a buffer between the gel-like clusters and the Ld
matrix where most of the unsaturated lipids are found, somewhat
resembling the role played by surfactant molecules in mixtures of oil
and water. The line tension between the two liquid phases may vanish
or not, depending on the affinity of the saturated and unsaturated
lipids which, in our simple model, is governed by the parameter
$\epsilon_{24}$. The MC simulations capture the transition from Type
II to Type I, taking place for $0.2<\epsilon_{24}<0.25$.

\backmatter


\bmhead{Acknowledgments}

This work was supported by the Israel Science Foundation (ISF), Grant
No. 1258/22. TS acknowledges the financial support from the Slovenian
Research Agency (research project No.~J1-3009 and research core
funding No.~P1-0055).

\section*{Declarations}


\begin{itemize}
\item Availability of data and materials

The data that support the findings of this study are available from
the corresponding author upon reasonable request.
  
\item Authors' contributions

TS and OF performed the research and wrote the article together.
\end{itemize}

\bibliography{ref.bib}

\end{document}